# Nucleotide String Indexing using Range Matching

Alon Rashelbach
Technion
Israel
alonrs@campus.technion.ac.il

Ori Rottensterich
Technion
Israel
or@technion.ac.il

Mark Silberstien
Technion
Israel
mark@ee.technion.ac.il

## ABSTRACT

The two most common data-structures for genome indexing, FM-indices and hash-tables, exhibit a fundamental trade-off between memory footprint and performance. We present Ranger, a new indexing technique for nucleotide sequences that is both memory efficient and fast. We observe that nucleotide sequences can be represented as integer ranges and leverage a range-matching algorithm based on neural networks to perform the lookup.

We prototype Ranger in software and integrate it into the popular Minimap2 tool. Ranger achieves almost identical end-to-end performance as the original Minimap2, while occupying 1.7× and 1.2× less memory for short- and long-reads, respectively. With a limited memory capacity, Ranger achieves up to 4.3× speedup for short reads compared to FM-Index, and up to 4.2× and 1.8× speedups for short- and long-reads, compared to hash-tables. Ranger opens up new opportunities in the context of hardware acceleration by reducing the memory footprint of long-seed indexes used in state-of-the-art alignment accelerators by up to 23× which results with 3× faster alignment and negligible accuracy degradation. Moreover, its worst case memory bandwidth and latency can be bounded in advance without the need to inflate DRAM capacity.

## 1 INTRODUCTION

High-throughput DNA sequencing technologies together with advances in sequence alignment tools have greatly accelerated biological research [4, 11, 19, 20, 30]. To perform the alignment, many tools use reference-guided assembly in which nucleotide sequences (*reads*) are aligned against a known reference genome. The alignment is commonly performed by hand-tuned heuristics. A common practice is to search for the most likely locations of read fragments within the genome [13, 15, 17]. The alignment outcome often carries some degree of inaccuracy due to sequencing errors.

An uncompressed reference genome index consumes tens of gigabytes of memory. Thus, the alignment process may require large DRAM capacity, which is known to be *one of the primary factors contributing to the total system cost* [1]. The memory cost is particularly high for the cases when the same alignment server is used by multiple users concurrently. While sometimes it is possible to reduce the required DRAM size by storing part of the index on an SSD and loading it on-demand (the mechanism called swapping), this approach results in a substantial slowdown due to the mostly random access pattern to the genome which precludes reusing it in memory.

The two most common data-structures for indexing the genome are FM-indices [7] and hash-tables [2]. They occupy two opposite points in the trade-off between performance and memory requirements. FM-indices are compact but slower due to their poor spatial locality [1]. In contrast, hash-tables are faster but occupy large amount of memory. To reduce the memory requirements, modern aligners use a very effective approach called *minimizers* [26]. In this work we show a method that, when used together with minimizers, allows reducing the memory footprint even further.

At the same time, the use of hardware accelerators in DNA processing pipelines is becoming essential to cope with exponential growth in the volume of sequenced DNA data [27]. However, state-of-the-art hardware accelerators either use *tens of GBs* for processing a *single reference genome* [5, 29], involve special memory technologies [3, 33], or require high memory bandwidth [32]. Thus, there is a clear need for alternative hardware-friendly, space-efficient and low-bandwidth index technique.

In this work, we present *Ranger*, a new technique for compact, hardware-friendly genome indexing that significantly improves the performance-vs-size trade-off. It is as fast as hash-tables used by state-of-the-art tools but occupies only *about a half* of their memory even when using minimizers. Ranger is easy to integrate with existing systems and does not require intrusive algorithmic modifications. To demonstrate that, we incorporate a software prototype of Ranger into the Minimap2 aligner [17] with minimal code changes.

**Results.** When considered for use with existing sequencing accelerators, Ranger enables dramatic reduction of the index size for longer seed sizes, thereby opening new opportunities for acceleration. Specifically, it makes the use of 18bp-long seeds practical in Darwin [29], reducing the index size from 270GB to 11GB (about 25× smaller). Compared to the 15bp-long seeds currently used in Darwin, the use of 18bp-long seeds improves the alignment rate by 3×, but without Ranger it would not have been cost-effective due to the prohibitively large memory footprint. Moreover, with Ranger, using 18bp-long seeds requires 2× less memory than using 15bp-long seeds in the original work [29].

We compare Ranger-enhanced Minimap2 along two axes: alignment performance and memory footprint. Ranger achieves almost identical end-to-end performance as unmodified Minimap2 while *occupying 42% less memory*. At the same time, in a memory-constrained environment Ranger demonstrates up to 4.2× and 1.8× faster end-to-end alignment rate for short- and long-reads respectively. Moreover, Ranger is up to 4.3× faster than BWA-MEM [15] while requiring 20% more memory.

**Key algorithmic ideas.** Ranger compresses the reference genome index by transforming it from a hash-table of key-value pairs (key: short nucleotide strings [2], value: location in the reference), into a map of *range* - set-of-locations pairs. Each range effectively represents multiple strings with the same prefix. Given a query string, Ranger uses a recent efficient *neural network-based algorithm for*

---

[1] https://memory.net/memory-prices/

[2] Nucleotide strings of a bounded length. See Section 3.0.2 for a formal definition.

*range matching*, RQRMI, [24] to find the matching prefix, and then search through the strings in the range to find the matching one and its locations within the reference.

Ranger is based on prior RQRMI work in the field of network packet processors. It demonstrates high cache locality and maps well to vector processing engines. Moreover, its worst case memory bandwidth and latency are bounded in advance without the need to inflate the DRAM footprint, in contrast to hash-tables, FM-indices, or position tables [9].

To summarize, our contributions are as follows. (1) We compress the reference genome index in memory by using ranges, and leverage a neural-net based range matching algorithm to query the compressed index. (2) We integrate Ranger into the popular Minimap2 aligner with minimal modifications to the original code. (3) We show that using Ranger, Minimap2 achieves superior end-to-end performance for machines with a limited memory capacity and similar performance otherwise, while occupying up to 1.7× less memory. (4) We show that Ranger has pros when used with hardware accelerators as it can shorten the processing time by up to 3× while reducing the index size by up to 25×. (5) We design and implement a software version of Ranger as a shared library, ready for integration with future prototypes.

## 2 BACKGROUND
### 2.1 Existing indexing techniques

Data-structures used for indexing a reference genome trade memory footprint for lookup speed. The two most common such data-structures are suffix-arrays and hash-tables [2]. Suffix-arrays such as FM-indexes [7] can perform queries of arbitrary input lengths and thus reduce the number of lookups [14, 18]. Moreover, their memory footprint is 1.5×-2× smaller than hash-tables. However, they suffer from excessive number of memory accesses and have poor spatial locality. Therefore, their performance is inferior to hash-tables when facing high error rates [1]. Improving performance is possible but at the expense of increasing the memory footprint by 10× [12, 31]. Hash-tables, on the other hand, perform queries on fixed-sized read fragments (*seeds*). They are, in fact, the most commonly used data structures for genome indexing [2]. Nonetheless, they too allocate extra memory to improve performance in order to reduce hash collisions and boost the lookup speed, e.g., 142% to achieve an average of 2.16 memory accesses per query [23].

### 2.2 Range matching

*Range matching* finds an integer interval, *range*, that contains an input integer out of a given dataset of ranges. For example, for the input integer 3 and the dataset $I_0 = [0, 4], I_1 = [5, 8]$, the algorithm produces $I_0$ as its output. In this work, we refer to the special case in which the ranges in the dataset do not overlap and the input domain is limited to 64-bit integers ($[0, 2^{64} - 1]$). Formally, given an ordered set of non-overlapping $R$ ranges $S = \{I_1, I_2, ..., I_R\}$ ($I_1 \cup ... \cup I_R = [0, 2^{64} - 1]$) and a 64-bit integer $x$, we say that the *matching range* of $x$ is $f(x) = l$ when $x \in I_l$ ($1 \le l \le R$). Traditional software techniques for range matching are either tree based or hash-table based. RQRMI [24] is a state-of-the-art range matching algorithm that uses neural networks (NNs) for learning the positions of the ranges within the dataset $S$. RQRMI uses a hierarchical neural-net model (described next) that may index millions of ranges yet is compact enough to fit the CPU first-level caches, thereby providing superior performance to existing alternatives.

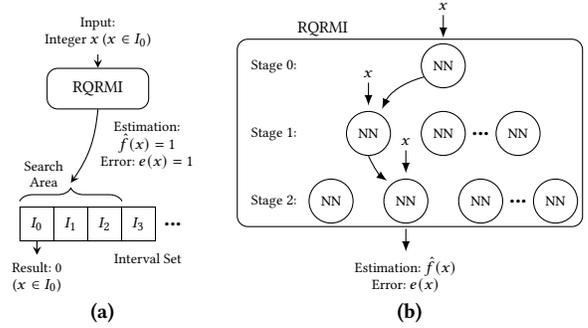

Figure 1: (a) Range matching with RQRMI. (b) RQRMI model structure.

### 2.3 The RQRMI model

The *Range-Query Recursive Model Index* (RQRMI) machine learning model learns the relative positions of the ranges in a sorted array that stores them. At the query time, RQRMI takes an integer as an input and outputs an *estimate* of the array index of the matching range. Note that the model is trained on ranges, but the query input is a single integer. The model is guaranteed to produce the index estimate within a predefined error bound computed at the training time.

Armed with the estimate of the index and the error bound, the query completes by performing a search over a subset of the ranges located within the error bound of the estimated index in the array. Formally, given an integer $x$, the RQRMI model outputs an estimate for the array index $\hat{f}(x)$ of the matching range (if it exists in the array) and an error $e(x)$ such that $\hat{f}(x) \in [f(x) - e(x), f(x) + e(x)]$ (Fig. 1a).

The RQRMI model consists of many shallow NNs organized in a hierarchy of at most three *stages*, as illustrated in Fig. 1b. The number of NNs per stage is predefined and depends on the number of ranges in the dataset, with the exception that the first stage must contain a single NN. An NN in an internal stage is used to select a single NN in the subsequent stage, while at each stage an NN is fed with the same input $x$. Thus, only one NN is evaluated per stage, with the total of at most three NN evaluations per query. Formally, denote the number of ranges as $R$ and the output of stage $i$ as $\hat{f}_i$. Given that stage $i + 1$ has $W$ NNs, the index of the NN in stage $i + 1$ that processes the input $x$ is $\lfloor \hat{f}_i(x)/R \cdot W \rfloor$. The last stage outputs the array index estimate $\hat{f}(x)$ and its error bound $e(x)$.

All NNs have the same structure: each consists of one fully-connected hidden layer with eight perceptrons and ReLU activation functions. Crucially, the RQRMI training algorithm guarantees a tight bound on the NNs' error which ensures lookup correctness and efficiency.

*2.3.1 Training the RQRMI model.* The RQRMI training process is performed stage-by-stage. First, it generates tuples of numbers $(x_1, ..., x_{32})$ and their locations within the array $(f(x_1), ..., f(x_{32}))$ by sampling the interval set. These tuples are used to perform supervised learning on the NNs such that the numbers are used as inputs and the locations as the labels to be learned. Once a stage is trained, its output is used for calculating the set of inputs that might reach each NN in the following stage. The training process then proceeds to the subsequent stage, and generates tuples per NN based on its calculated set of inputs. This process continues for all stages. After all NNs are trained, the model error is calculated, as described in [24]. If the error is larger than a predefined threshold, the training process repeats for another round, and stops after a predefined number of trials.

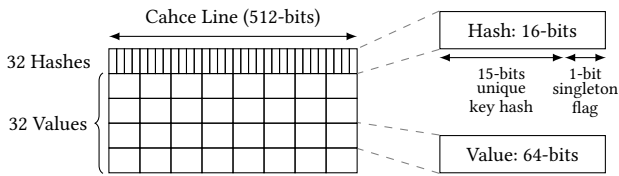

Figure 2: The Ranger bucket layout when using 64-bit values.

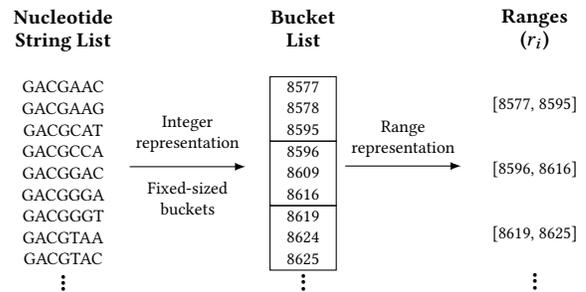

Figure 3: Conversion from sorted nucleotide strings to integer ranges.

*2.3.2 RQRMI performance benefits.* The RQRMI model may index millions of ranges with a relatively small error while requiring about 64KB of memory. Such models are small enough to fit the L1/L2 CPU caches, thus their use for range matching turns the task into *compute bound*. The small model size translates into a performance advantage compared to hash-tables or trees because it saves costly DRAM memory accesses.

## 3 RANGER DESIGN AND IMPLEMENTATION

We propose Ranger, a new technique for indexing nucleotide strings using range matching. In contrary to previous methods, its unique combination of high performance and low memory footprint makes it suitable for systems with low RAM budgets as well as for genome hardware accelerators. We now describe its design and implementation principles.

*3.0.1 Overview.* Ranger takes a lexicographically sorted list of unique short nucleotide strings of up to 32 nucleotides (keys) and their positions within the reference genome (values) and divides it into fixed-sized buckets. The strings that fall into the same bucket have the same prefix. Thus, each bucket can be represented as an integer range, as we discuss in Section 3.0.2, which makes it possible to use range matching as part of the bucket lookup process.

The bucket structure is illustrated in Fig. 2. To save space, nucleotide strings are not stored explicitly in the buckets. Rather, a bucket stores 15-bit hash of each string. The bucket contains up to 32 hashes of the strings and their corresponding positions in the reference genome. Since we store hashes false positive matches are possible. However, the overall read mapping accuracy impact is negligible, as we show in Section 3.0.4.

A bucket holds a single value for each string. If a string is a singleton (i.e., there is only a single position in the reference genome), this position is stored in the bucket. But if there are multiple positions, such *non-singleton entries* hold a pointer to their complete position list. The combined position list for all non-singleton entries is stored separately from the buckets.

Ranger can be configured to either hold 64- or 32-bits per value, depending on the aligner requirements. For example, Minimap and Minimap2 aligners use 64-bit values for storing the target sequence index, the position of the Minimizer, and additional strand information [16].

A query is performed as follows. Given a nucleotide string $t$, an RQRMI model finds the matching range which corresponds to a bucket $b$ that contains $t$. Given $b$, we search for $hash(t)$ in $b$. If a match is found, the corresponding value is returned to the user if $t$ is a singleton. Otherwise, we fetch all the respective positions from the combined position array.

**Hardware-friendly implementation.** Buckets are specifically built to enable efficient CPU implementation. Specifically, all the string hashes are stored in a single CPU cache line. They can be checked in parallel using vector instructions. Each bucket is aligned to a cache line, i.e., it occupies several full cache lines to ensure access efficiency to the keys.

**RQRMI configuration.** Ranger uses a small RQRMI model with three stages for indexing the ranges. The number of NNs per stage is determined by the number of ranges the model index, with a configuration of {1, 8, 119} NNs that correspond to 16KB total memory footprint. The configuration was chosen empirically so that the RQRMI error would be 1024 in the worst-case scenario. On average, the model error is 44 elements which corresponds to a binary search over six cache lines, and requires at most three memory accesses per lookup.

**Space requirements.** The Ranger index data structure consists of the RQRMI model (16KB), an array of ranges (64-bit per range, see Section 3.0.2), an array of buckets (320B per bucket using 64-bit values), and an array of values for the non-singleton entries. Apart from a few underutilized buckets (Section 3.0.4), the Ranger index data structure is dense. Thus, it is more space-efficient than a hashtable as it does not require allocating additional space to reduce collisions.

*3.0.2 Representing string buckets as ranges.* Let $\Sigma=\{A,C,G,T\}$ be the alphabet of nucleotides. In the following, we refer to each of these characters as *base-pairs* (bp). We consider a nucleotide string $s \in \Sigma^*$ as *short* if it contains at most 32 bp ($|s| \leq 32$). By using two bits for encoding each letter in $\Sigma$, a lexicographically sorted list of short nucleotide strings $L$ can be represented as a sorted list of 64-bit integers. This list can is divided into buckets with a fixed number of integers, as mentioned in Section 3.0.1. The ordered set of integers $\{i_1, i_2, ..., i_n\}$ within a bucket define a range $r = [i_1, i_n]$ for that bucket. See Fig. 3 for illustration. We discuss the case in which $L$ contains *unique* nucleotide strings; this is possible since we regard identical strings as a non-singleton entries (see Section 3.0.1). Let $r_l = [b_l, e_l], r_k = [b_k, e_k]$ be two ranges generated from $L$ that correspond to buckets $l$ and $k$ respectively. From the uniqueness of elements in $L$, if $b_l < b_k$ then $e_l < b_k$, meaning the ranges do not overlap. In this case, we say that $r_l < r_k$.

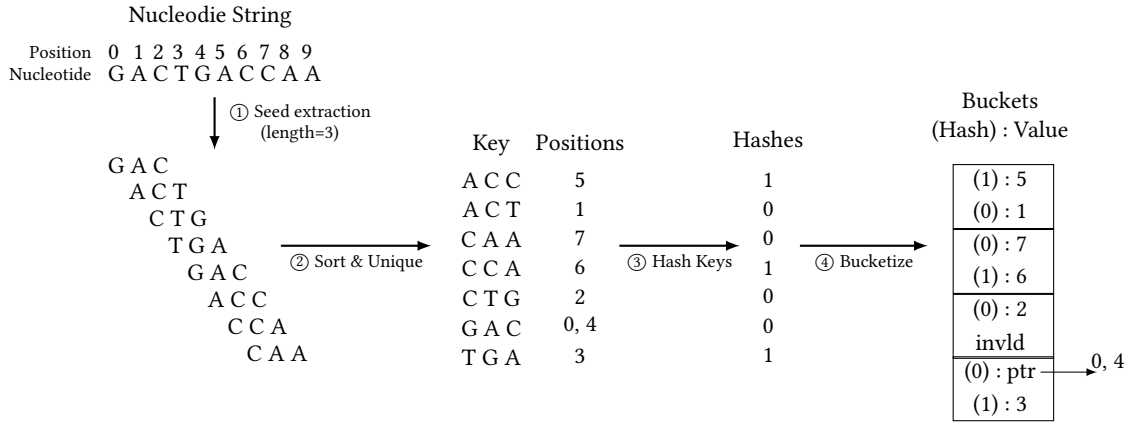

**Figure 4: Bucket generation in Ranger.** ① Short substrings are extracted from the reference genome using any arbitrary method, e.g., Minimizers ② The substrings are sorted and duplicates are marked as non-singleton entries. ③ Keys are hashed and buckets are generated such that they do not contain duplicate hashes. Duplicate hashes are put in different buckets. ④ Underutilized buckets contain invalid entries. Non-singleton entries point to a separate position list.

For each range, we store only its lower bound to save memory. This is sufficient to produce a correct match because ranges do not overlap and the algorithm always scans the resulting bucket for the matching string. This optimization comes at the cost of the lookup performance, as the bucket scanning time is the slowest component (see Section 4.2.5).

*3.0.3 Bucket generation.* Ranger builds buckets one-by-one by iterating over the sorted list of nucleotide strings and populating their hashes and values within the buckets, as illustrated in Fig. 4. Whenever it encounters a nucleotide string that causes hash collision in bucket $i$, it marks the remaining entries in that bucket as invalid and continues to bucket $i+1$. This is essential to avoid ambiguity in the lookup process by ensuring a single entry per hash match. Although buckets might not reach full utilization, underutilized buckets have insignificant effect on the memory footprint (see Table 1).

*3.0.4 Mapping accuracy and false positives.* Ranger uses 15-bits for storing string hashes and marks invalid entries with zero hash values. Since there are at most 32 hashes per bucket, false positive matches occur with probability $32 \cdot (2^{15} - 1)^{-1} < 0.1\%$. We test the mapping accuracy by simulating short and long reads using the

**Table 1: Various statistics for the Ranger index data structure for both 32-bit and 64-bit value size configurations.**

| Reads | Keys | Singles | Buckets | Util. | Size (GB) 32-bit | 64-bit |
|---|---|---|---|---|---|---|
| Short | 493M | 96.7% | 12.9M | 99.6% | 2.79 | 4.77 |
| Long | 545M | 38.7% | 3.1M | 99.9% | 2.55 | 4.91 |

*Notes:* The two data structures index the human reference genome GRCh38 using the Ranger plugin for Minimap2 (see Section 3.0.5). The short read preset uses (28,11)-Minimizers and the long read preset uses (15,10)-Minimizers. A $(k, w)$-Minimizer is the smallest $k$-mer in a window of $w$ consecutive $k$-mers [26].

methodology described in Section 4.1. The results in Fig. 5 demonstrate that the effect of false-positives matches on the mapping accuracy is negligible.

*3.0.5 Implementation.* Ranger is implemented using C++ as a dynamic library with a simple API compatible with C applications. To achieve maximum performance, it incorporates AVX2 SIMD instructions which are available in Intel CPUs since 2012. The library (*libranger*) has no external dependencies except the library for fast RQRMI training (*libnmu*), provided as part of the Ranger repository [25].

We integrate Ranger into the state-of-the-art sequence aligner Minimap2[3] [17]. We select Minimap2 for its accuracy and relative performance over other popular aligners for both short and long reads. Overall, the integration contains 840 new lines of code, out of which 93 replace original code. Specifically, we had to alter the Minimap2 code to support batched queries to be compatible with Ranger. Note that Ranger is oblivious to Minimap2 internals and is only responsible for generating the index data-structure and performing queries on it. Specifically, Ranger is unaware of the fact that Minimap2 works with Minimizers; nevertheless, we found a Minimizer configuration for short reads that works better for Minimap2 with Ranger (see Table 1). The documentation of *libranger* alongside its code and the Minimap2 integration are freely available at https://github.com/acsl-technion/ranger.

---
[3]We use Minimap2 version 2.22-r1105 committed on 16-Aug-2021. https://github.com/lh3/minimap2/tree/05a8a4

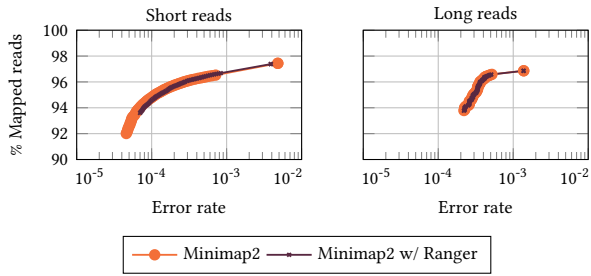

**Figure 5: Mapping accuracy of simulated reads using Minimap2 with and without Ranger.** Storing hashes of strings does not have visible effects on the mapping precision. The accuracy measurement methodology is similar to the one presented in the Minimap2 paper. A read is considered correctly mapped if its longest alignment overlaps at least 10% of its ground truth position within the reference. We sort the resulting mapping qualities in descending order, then for each mapping quality threshold we plot the percentage of aligned reads with mapping quality above the threshold and their error rate. See Section 4.1 for more details on the simulated datasets.

## 4 SOFTWARE PROTOTYPE RESULTS

We evaluate the impact of Ranger software prototype on the end-to-end performance of the popular aligner Minimap2. Our primary question is how the size reduction achievable using Ranger affects systems with a limited RAM budget, and whether the effect scales with the number of cores. We also show how Ranger behaves compared to hash-tables when the memory capacity is not the primary bottleneck.

### 4.1 Methodology

All experiments were performed on a Linux machine with Ubuntu 18.04 (Linux kernel 5.4) and an Intel Xeon Silver 4116 CPU @ 2.1 GHz with 32KB L1 cache, 1024KB L2 cache, and 16.5MB LLC. We disable hyper-threading and allocate one CPU core per software thread. We use Linux cgroups to limit the RAM capacity for simulating various machine memory configurations. Unless stated otherwise, experiments were performed with eight software threads. We test various aspects of Ranger integrated into Minimap2. For the baseline we use Minimap2 with batched queries, which is slightly faster than the original Minimap2 version. We set Ranger to use 64-bit values as also used in the original Minimap2 for the sake of a fair comparison. We use the default Minimap2 configurations '–x sr' and '–x map-ont' for aligning short and long reads, respectively. Ranger uses minimizer sizes for short reads that differ from Minimap2 defaults, with (28,11)-minimizers and (15,10)-minimizers for short- and long-reads respectively, compared to (21,11)-minimizers and (15,10)-minimizers in the original Minimap2. These configurations were empirically selected due to their superior results in both methods. Short reads alignment performance is also compared against the popular BWA-MEM aligner with default configuration [15].

*4.1.1 Datasets.* Our evaluation uses both synthetic and real world sequencing data. Real data for short reads was obtained from EMBL-ENA accessories ERR194147_1 (787M reads of length 101bp from which we randomly sampled 100M reads), SRR826460_1 (89M reads of length 150bp), and SRR826471_1 (186M reads of length 250bp). For long reads, we use 668K MinION reads with N50 of 11,218 [11]. We also simulate 10M short reads of length 150bp and 200K long reads with N50 of 12,716bp using Mason2 [10] and PBSIM2 [22], respectively. For Mason2, we use the '–illumina-prob-mismatch-scale 2.5' option, similar to the setting reported in the Minimap2 paper. For PBSIM2, we use the PacBio P6-C4 chemistry. Index data structures were generated on the human reference genome GRCh38 [6].

*4.1.2 Accuracy measurements.* Our measurement of the end-to-end mapping accuracy is similar to the one presented in the Minimap2 paper. A read is considered correctly mapped if its longest alignment overlaps at least 10% of its ground truth position within the reference. We sort the resulting mapping qualities in descending order, then for each mapping quality threshold we plot the percentage of aligned reads with mapping quality above the threshold and their error rate.

### 4.2 End-to-end results

*4.2.1 Performance.* We test Ranger's end-to-end performance on real data as a function of the memory capacity. The results illustrated in Fig. 6a and Fig. 6b show that Ranger is up to 4.2× faster (1.7× geometric mean) than the original Minimap2 for machines with a limited memory budget and that it *achieves roughly the same performance as the baseline using 1.5× less RAM* (for example, the performance of vanilla Minimap2 in a 12GB server are identical to the performance of Minimap2 with Ranger in a 8GB server). In addition, Ranger achieves between 1.9×-4.2× (2.7× geometric mean) speedup compared to BWA-MEM, while occupying 1.2× more memory footprint. When the memory is not the primary bottleneck, Ranger's performance is similar to the baseline for three out of the four data-sets. The only data-set in which Ranger is worse is the 100M 101bp data-set, with a 5Min (16%) slowdown. This behavior originates from the difference in the minimizer configuration between the two methods, while using the same minimizer sizes yields similar performance results. Still, we believe that the (28-11)-minimizer configuration is more suitable for Ranger as it demonstrates better results for 14 out of 15 experiments.

*4.2.2 Memory footprint.* Table 2 presents the breakdown of Minimap2 memory footprint when using hash-tables and Ranger as its indexing techniques respectively, with minimizers enabled. Overall, the Minimap2 memory footprint while using Ranger is smaller by 1.72× and 1.18× for short- and long-reads, respectively. Both methods need to store the reference genome and the position lists of non-singleton entries alongside their index. Consequently, any difference in the total memory footprint between the two originates from the density of Ranger's data-structure, which increases with

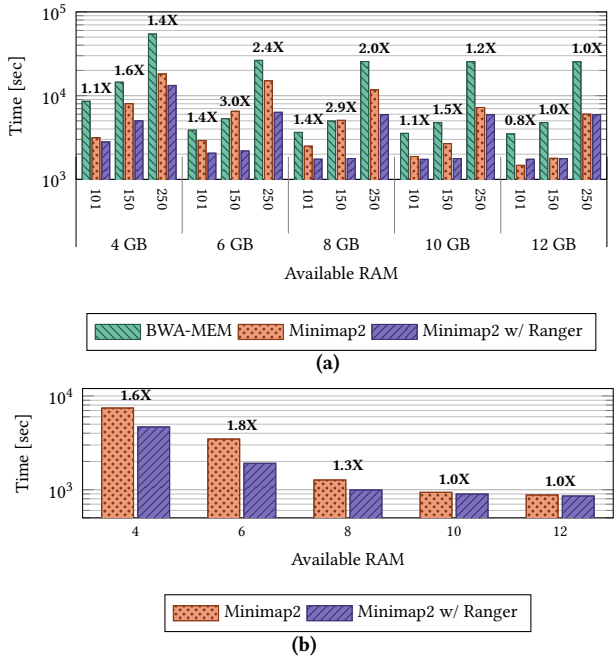

(a)

(b)

Figure 6: Sequence alignment time with and without Ranger for short- and long-reads of five RAM size categories (lower is better). See Section 4.1.1 for more information about the evaluated datasets. The performance speedups of Ranger compared to the original Minimap2 tool are annotated in bold text above the bars. (a) Results for short-reads. The x-axis labels 101, 150, and 250 correspond to short reads with 101, 150, and 250 base-pairs, respectively. (b) Results for long-reads.

Table 2: The index size in MB using hash-tables and Ranger, both with 64-bit values.

|  | Short Reads | | | Long Reads | | |
| --- | --- | --- | --- | --- | --- | --- |
|  | Hash Tables | Ranger | × | Hash Tables | Ranger | × |
| Idx | 8321 | 4051 | 2.0× | 2082 | 980 | 2.1× |
| Pos | 981 | 719 | 1.4× | 3937 | 3937 | 1× |
| Str | 1478 | 1478 | 1× | 1478 | 1478 | 1× |
| Total | 10780 | 6248 | 1.7× | 7498 | 6395 | 1.2× |

*Notes:* We used the human reference genome GRCh38 and Minimap2 with Ranger. *Idx* stands for the index data-structure, e.g., the hash-table or the bucket array. *Pos* is the combined position list of all non-singleton entries. *Str* stands for the reference nucleotide string.

the number of unique bucket entries. Moreover, the percentage of the position lists out of the overall memory footprint diminishes as the number of singletons increases (e.g., as in the case of short reads), which further helps reduce the index size.

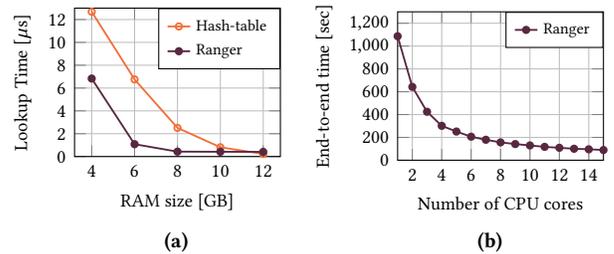

(a)

(b)

Figure 7: (a) The effect of the memory capacity on the average query time using Minimap2 with hash-tables and with Ranger. The curves represent the geometric mean over all datasets. (b) Minimap2 end-to-end alignment time using Ranger for short reads as a function of the number of cores.

Table 3: Breakdown of the Ranger query time to its basic components.

|  | RQRMI Inference | Range Array Search | Bucket Lookup |
| --- | --- | --- | --- |
| Time [ns] | 45 | 218 | 337 |
| Percentage (%) | 8 | 36 | 56 |

*4.2.3 The source of speedups.* Minimap2 uses an efficient hash-table implementation that uses extra-memory to avoid hash collisions. In contrast, Ranger can occupy half the memory without performance penalties. As such, limiting the memory capacity causes the hash-table to spill out of the main memory into slow storage and suffer from storage accesses as part of the lookup process. As server memory capacity decreases, storage accesses become the dominating factor of the lookup process, as illustrated in Fig. 7a.

*4.2.4 Scalability with the number of CPU cores.* We evaluate the scalability of Ranger with the number of cores by measuring its end-to-end performance on simulated short reads. We use 16GB of RAM to keep the reference genome in memory. The results in Fig. 7b show a diminishing return in performance as the number of cores increases. This behavior is not unique to Ranger as it is also observed in the baseline. Both methods achieve nearly identical results. This indicates that Ranger does not induce additional overheads when using multiple cores.

*4.2.5 Understanding Ranger's lookup latency.* Table 3 shows the breakdown of Ranger's query time to its basic components: RQRMI inference, binary search over the range array, and the bucket internal lookup. We report the performance results of the short read index. With a 56% of the total latency, the bucket internal lookup is the slowest. Given the large memory footprint of the bucket array and its random access nature, bucket lookups often result with minor page faults (i.e., TLB misses) that hinder the lookup performance. Huge-tables can help reduce the effect of TLB misses. The binary search occupies 36% of the total lookup latency. The range array of the short read index occupies 98.4MB (12.9M ranges, 8B per range, see Table 1), so it does not fit the CPU cache and the RAM access latency governs the search performance. The RQRMI

Nucleotide String Indexing using Range Matching

worst-case error is 1024, hence the binary search takes at most 10 memory accesses. Yet, the average case requires three memory accesses. The fastest component is the RQRMI inference, with 8% of the total lookup time. This is expected, as the RQRMI model is designed to fit within the CPU's L1 cache and to use efficient SIMD instructions.

*4.2.6 Ranger vs. FM-index.* We did not conduct a direct comparison between the two due to the difficulty of decoupling the indexing technique from other aspects of the alignment process. FM-indices are known to be significantly slower than hash-tables [2], partly for their inability to take advantage of CPU characteristics such as cache line access granularity or memory prefetching [31]. As an example, the Minimap2 tool uses an efficient hash-table implementation and achieves a 3×-4× and 30× faster end-to-end alignment speeds compared to the FM-index-based Bowtie2 [13] and BWA-MEM [15] for short- and long-reads, respectively. Since Ranger achieves a comparable end-to-end alignment performance over hash-tables (Section 4.2), it is safe to assume that it is faster than FM-indices. When it comes to memory footprint, it depends on Ranger's configuration: for the 64-bit value size setting Ranger is 30% larger than the equivalent FM-index and for the 32-bit setting their size is roughly the same [4].

## 5 HARDWARE ACCELERATION BENEFITS

Ranger can benefit hardware accelerators for sequence alignment [8, 21, 28, 29]. State-of-the-art methods for long-read alignment focus on hash-tables or seed position tables [9] for their efficiency compared to suffix-trees or FM-indices. However, the trade-offs exhibited by both hash-tables and seed position tables require compromises between performance and memory capacity. For example, Drawin [29], a state-of-the-art hardware accelerator for long-read alignment, uses a seed position table for indexing seeds of 15bp length. However, the use of such a small seed length is suboptimal from the performance perspective. Our experiments show, for example, that 18bp seeds improve the end-to-end alignment performance by 3× (Fig. 8a) thanks to the high percentage of singleton seeds (Fig. 8b) while having a minor accuracy impact (Fig. 8c). Yet, using seed position tables for large seeds is impractical due to their large index size, e.g., 272GB for 18bp seeds compared to 20GB for 15bp seeds (Fig. 8d). Hash-tables are smaller but not by much: their total index size is 75.5GB for 18bp seeds. In comparison, Ranger can index 18bp seeds with only 11.7GB while using the 32-bit value setting.

Another advantage of Ranger is its memory bandwidth and latency: they are deterministic and bounded, making the hardware implementation much simpler. The small size of the RQRMI model makes it fit the fast on-chip memory and its inference process can be efficiently pipelined as it uses a constant number of floating-point operations. The range array can be transformed to fit the fast on-chip memory so the binary search can be performed efficiently without accessing the DRAM. This transformation has been described in [25]. In essence, adjacent ranges in the range array are merged to form a single large range, effectively reducing the number of ranges to be stored in the range array indexed by the RQRMI model. This smaller array is stored in the fast on-chip memory. The original ranges are stored in DRAM off-chip. After RQRMI determines the large range, the algorithm fetches the constituent original ranges from DRAM.

---

[4]We refer to the compressed version of the FM-index that occupies 4.7GB, not to the uncompressed version which occupies 48GB [31].

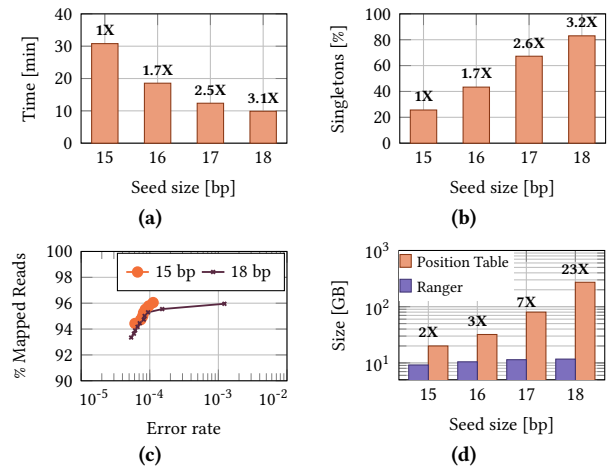

(a)

(b)

(c)

(d)

**Figure 8: The implications of using different seed lengths for performing long read alignment. The experiments were performed using Minimap2 on simulated long reads (see Section 4.1). (a) The end-to-end processing time decreases as the seeds' length increases. The bold annotations represent performance speedups compared to 15bp. (b) The percentage of singleton entries increases as the seeds' length increases. The bold annotations represent the ratio compared to 15bp. (c) The impact of the end-to-end mapping accuracy as a function of the seeds' lengths. Seed lengths of 16 and 17bp behave similarly. See Section 4.1 for the methodology of the accuracy measurement. (d) The memory footprint of both the seed position table index data structure [9] and Ranger. The bold annotations represent the ratio between the two.**

For example, by merging together eight buckets from the short-read preset (see Table 1), the new range array occupies 12.25MB and can fit the fast on-chip memory of modern ASICs. Following the binary search, a single DRAM access fetches the constituent ranges so the appropriate bucket could be found. In this example, this technique requires one DRAM access of 64B per lookup. Fetching the appropriate bucket requires an additional DRAM access. For 32-bit values, the algorithm fetches an additional 192B per lookup. In cases where non-singleton entries are matched, another DRAM access is required for fetching the appropriate location lists. Using the common optimization of discarding high-frequency seeds makes this access deterministic in terms of bandwidth. In our example, discarding seeds with more than 64 locations would result with another 256B per lookup. But high-frequency seeds are quite rare because Ranger uses large seed lengths, resulting in 17% for 18bp seeds in our example.

To conclude, a hardware implementation of Ranger requires three DRAM accesses and 512B of DRAM bandwidth per lookup in the worst-case scenario, and 2.17 accesses of 300B in the average-case. This result corroborates the benefits of Ranger for indexing reference genome in hardware accelerators.

## 6 CONCLUSION

We have presented Ranger, a new technique for nucleotide string indexing using range matching. Ranger takes advantage of RQRMI models to perform fast range matching as part of the seed lookup procedure, thus reducing the size of the index data-structure to half of the equivalent hash-table. Ranger is not limited by the performance-vs-size trade-off as exhibited in both FM-indices and hash-tables. Compared to the state-of-the-art Minimap2 tool, Ranger achieves up to 4.2× end-to-end speedup for machines with a limited memory capacity. When the memory is not the primary bottleneck, Ranger demonstrates a performance similar to hash-tables while occupying up to 42% less memory footprint. Ultimately, its limited memory bandwidth and predictable latency makes it suitable for next-generation hardware accelerators for sequence alignment.